\documentclass[twocolumn, deluxetables]{aastex631}
\usepackage{amsmath}
\usepackage{multirow}
\usepackage{graphicx}
\graphicspath{{./}{figures/}}

\shorttitle{NGC 5033}
\shortauthors{Yun et al.}

\begin{document}

\title{Extreme X-ray Reflection in the Nucleus of the Seyfert Galaxy NGC 5033}

\correspondingauthor{J. M. Miller}
\email{jonmm@umich.edu}

\author{S.~B.~Yun}
\author{J.~M.~Miller}
\affil{Department of Astronomy, The University of Michigan, 1085 S. University Ave., Ann Arbor, MI, 48109, USA}
\author{D.~Barret}
\affil{IRAP, Universite de Toulouse, CNRS, UPS, CNES 9, Avenue du Colonel Roche, BP 44346, F-31028, Toulouse Cedex 4, France}
\author{D. Stern}
\affil{Jet Propulsion Laboratory, California Institute of Technology, 4800 Oak Grove Drive, Pasadena, CA 91109, USA}
\author{W. N. Brandt}
\affil{Department of Astronomy \& Astrophysics, The Pennsylvania State University, 525 Davey Laboratory, University Park, PA, 16802, USA}
\affil{Center for Gravitation and the Cosmos, The Pennsylvania State University, University Park, PA, 16802, USA}
\affil{Department of Physics, The Pennsylvania State University, 104 Davey Laboratory, University Park, PA, 16802, USA}
\author{L. Brenneman} 
\affil{Center for Astrophysics, Harvard \& Smithsonian, 60 Garden Street, Cambridge, MA 02138, USA}
\author{P. Draghis}
\affil{Department of Astronomy, The University of Michigan, 1085 S. University Ave., Ann Arbor, MI, 48109, USA}
\author{A. C. Fabian}
\affil{Institute of Astronomy, University of Cambridge, Madingley Road, Cambridge CB3 OHA, United Kingdom}
\author{J. Raymond}
\affil{Center for Astrophysics, Harvard \& Smithsonian, 60 Garden Street, Cambidge, MA 02138, USA}
\author{A. Zoghbi}
\affil{Department of Astronomy, The University of Michigan, 1085 S. University Ave., Ann Arbor, MI, 48109, USA}




\begin{abstract}
NGC 5033 is an intriguing Seyfert galaxy because its sub-classification may change with time, and because optical and sub-mm observations find that the massive black hole does not sit at the dynamical center of the galaxy, pointing to a past merger.  We obtained a new optical spectrum of NGC 5033 using the 200'' Hale telescope at Palomar that clearly reveals a broad H$\beta$ line (FWHM$=5400\pm 300~{\rm km}~{\rm s}^{-1}$). This signals a clear view of the optical broad line region (BLR) and requires Seyfert-1.5 designation.  Some spectra obtained in the past suggest a Seyfert-1.9 classification, potentially signaling a variable or "changing-look" geometry.  Our analysis of a 2019 Chandra spectrum of the massive black hole reveals very little obscuration, also suggesting a clean view of the central engine.  However, the narrow Fe~K$\alpha$ emission line is measured to have an equivalent with of EW$=460^{+100}_{-90}$~eV.  This value is extremely high compared to typical values in unobscured AGN.  Indeed, the line is persistently strong in NGC 5033: the line equivalent width in a 2002 XMM-Newton snapshot is EW$=250^{+40}_{-40}$~eV, similar to the EW$=290^{+100}_{-100}$~eV equivalent width measured using ASCA in 1999.   These results can likely be explained through a combination of an unusually high covering factor for reflection, and fluxes that are seen out of phase owing to light travel times.  We examine the possibility that NGC 5033 may strengthen evidence for the X-ray Baldwin effect.

\end{abstract}



\section{Introduction} \label{sec:intro}

NGC 5033 is a nearby barred spiral galaxy, at a redshift of $z=0.0029$ and a distance of 16.2 Mpc (Springbob et al.\ 2005).  The galaxy is viewed at a relatively high inclination angle, but dust lanes do not appear to lie along the line of sight to the galactic center (Pogge \& Martini 2002).  The active nucleus in NGC 5033 was studied as part of the CfA Seyfert Sample (Huchra \& Burg 1992); soon thereafter, it was classified as a Seyfert 1.9 (Osterbrock \& Martel 1993).  It has also been considered in the context of ``unobscured'' Seyfert-2 AGN (e.g., Panessa \& Bassani 2002), although as a source that differs from more typical examples of this sub-class.  
Ho et al.\ (1997) classified NGC 5033 as a Seyfert-1.5, while Veron-Cetty \& Veron (2006) classified it as a Seyfert 1.8.  It is possible that all of these classifications are correct, and that the appearance of NGC 5033 changes with time.  

From the perspective of understanding the co-evolution of black holes and host galaxies, the most interesting feature of NGC 5033 is that the black hole may not sit at the dynamical center of the galaxy.  The center of the H$\beta$ rotation is 3'' away from the center of the continuum emission around the H$\beta$ line (Mediavilla et al.\ 2005).   Moreover, the CO profile of the nucleus in NGC 5033 appears to be disturbed, and to potentially have two peaks (Kohno et al.\ 2003).  In concert, these findings suggest that NGC 5033 underwent a recent merger.  Relative to other galaxies, then, it is possible that the accretion flow -- on scales of the molecular torus and larger -- may not have settled into a standard configuration in NGC 5033.

X-rays are an excellent means of probing the accretion flows onto massive black holes, owing to their ability to penetrate gas and dust.  A Chandra observation made in 2015 (ObsID 16354) was plagued by strong photon pile-up (Davis 2001), distorting the image of the nuclear region and burying X-ray lines in a falsely elevated continuum spectrum (e.g., Miller et al.\ 2010).  One of two archival XMM-Newton snapshots was ruined by soft proton flaring.  The best prior observation and analysis of the X-ray spectrum of NGC 5033 may come from a 39~ks ASCA exposure (Terashima et al.\ 1999).  That observation revealed an unusually strong and spectrally unresolved Fe~K$\alpha$ emission line, with an equivalent width of EW$=290\pm100$~eV.  The source luminosity was also found to be low relative to most famous Seyferts (see, e.g., Nandra et al.\ 2007, Shu et al.\ 2010), $L = 2.3\times 10^{41}~{\rm erg}~{\rm s}^{-1}$ (2--10~keV).  Over the 0.5-10.0 keV fitting band, a very low line-of-sight column density of just $N_{H} = 8.7\pm 1.7 \times 10^{20}~{\rm cm}^{-2}$ was measured, consistent with an unobscured AGN.

In AGN and other systems, neutral Fe~K$\alpha$ lines result from the irradiation of cold gas by hard X-ray emission, a process known as ``reflection'' (e.g., George \& Fabian 1991).  Given (1) a fixed viewing angle, (2) gas that remains cold, and (3) a steady irradiating flux, the strength of the Fe~K$\alpha$ line encodes how much of the sky is covered by the reflecting gas.  Subject to these caveats, the ``reflection fraction'' is defined as $R = \Omega/2\pi$ within prevalent models (where $\Omega$ is the solid angle of the reflector as viewed from the central source).  Especially if the line is {\em narrow}, it is likely to be produced at a large distance from the black hole.  For this reason, {\em narrow} Fe~K$\alpha$ emission lines are often associated with the distant ``torus'' in AGN unification schemes (e.g., Antonucci 1993).  Recent work suggests that some of these lines are not truly narrow and likely originate within the optical broad line region (BLR), or even closer to the black hole (Shu et al.\ 2010, Miller et al.\ 2018, Zoghbi et al.\ 2019).  It is also clear that Fe~K$\alpha$ line flux can originate on kpc scales (e.g., Ponti et al. 2010, Fabbiano et al. 2017, Marinucci et al. 2017, Jones et al. 2020, Yan et al. 2021).

We requested a new Chandra imaging observation of NGC 5033 in Cycle 20 in order to separate the nuclear source from any surrounding point sources or diffuse emission.  The modest spectral resolution of the ACIS array does not typically permit very strong constraints on the width of X-ray lines; however, this is partially mitigated when a line is very strong.  For an additional constraint on the evolution of the nuclear source spectrum with time, we analyzed an archival XMM-Newton snapshot observation, and obtained an optical spectrum using Palomar.  These observations and our data reduction methods are detailed in Section 2.  In Section 3, we present our analysis and results, including the discussion of NGC 5033 in the context of the ``X-ray Baldwin effect'' (Iwasawa \& Taniguchi 1993). The implications of our work and paths to better understand the accretion flow onto the black hole in NGC 5033 are presented in Section 4.

\begin{figure*}[t]
\centering
\includegraphics[scale=0.6]{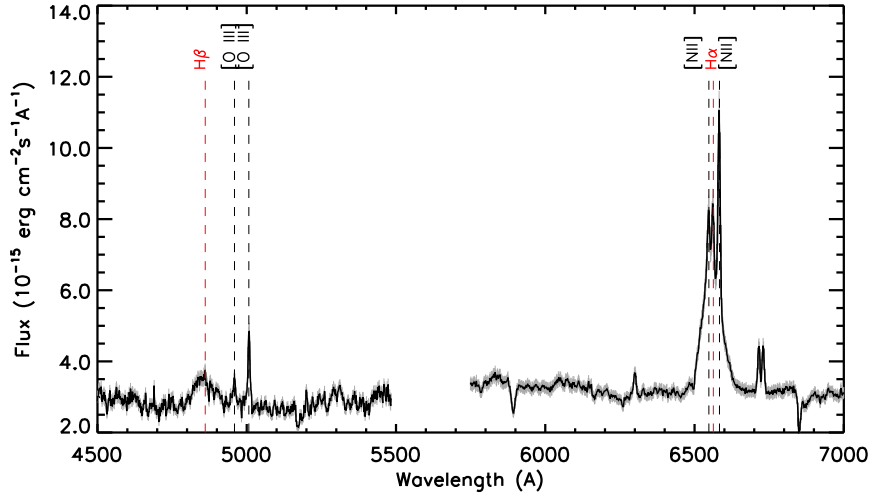}
\vspace {1mm}
\caption{Palomar optical spectra of NGC 5033.  Broad H$\alpha$ (FWHM$=3400\pm 50~{\rm km}~{\rm s}^{-1}$) and H$\beta$ lines (FWHM$=5400\pm 300~{\rm km}~{\rm s}^{-1}$) signal that the central engine is not entirely obscured, and point to a Seyfert-1.5 classification.  This strongly implies that the strength of the narrow Fe~K$\alpha$ line and the overall X-ray spectrum cannot be explained through a Compton-thick AGN and scattered emission.}
\label{fig:opt}
\end{figure*}

\section{Observations and Data Reduction}
\subsection{Palomar}

We observed NGC~5033 on UT 2021 September 3 from the Hale 200'' telescope at Palomar Observatory using the dual-beam, optical Double Spectrograph (DBSP). The night was photometric with 1\farcs1 - 1\farcs4 seeing. We obtained a single 300~s observation using the 1\farcs5 slit, the 5600~\AA\, dichroic, the 600 $\ell\, {\rm mm}^{-1}$ blue grism ($\lambda_{\rm blaze}$ = 4000 \AA), and the 400 $\ell\, {\rm mm}^{-1}$ red grating ($\lambda_{\rm blaze}$ = 8500 \AA). This instrument configuration covers the full optical window at moderate resolving power, $R \approx 1000$, with a modest gap at the dichroic. Flux calibration was obtained using observations of the subdwarf O-star HZ44 and the DA0 white dwarf G191-B2B from Massey \& Gronwall (1990) obtained on the same night. The spectra were reduced using standard IRAF routines.  

\subsection{Chandra}

\begin{figure*}[th]
\centering
\includegraphics[scale=0.58]{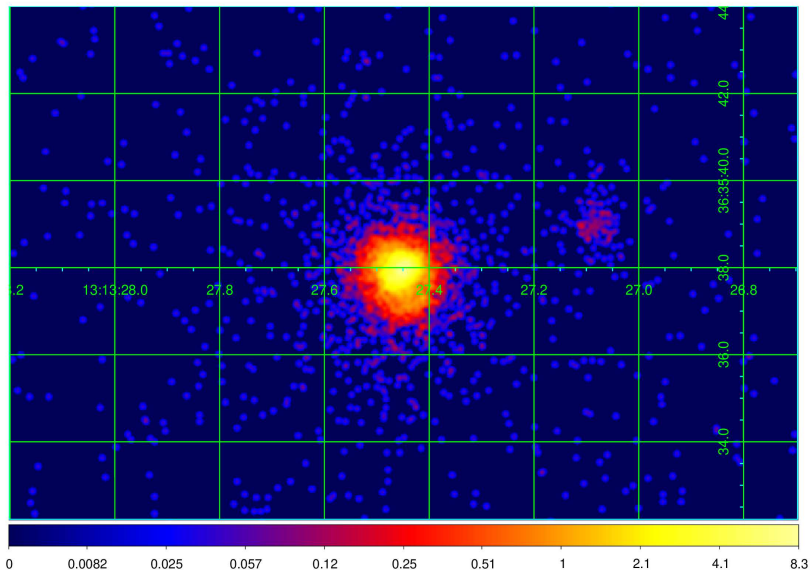}
\vspace {1mm}
\caption{The Chandra ACIS/S3 image of the nucleus of NGC 5033 in the 0.3-8.0 keV band.  Sub-pixels representing 0.1 native pixels were created, and then smoothed by a factor of 3.0 in order to create this image.   The nucleus is at the center of the field.  A weak neighboring source, also present in prior Chandra exposures, may be marginally resolved (0.3-0.4 arc seconds in extent) but physical arguments suggest that it is more likely a bright X-ray binary, potentially just above the ULX threshold.  The color bar indicates the pixel value.}
\label{fig:chandra_img}
\end{figure*}

Chandra observed NGC 5033 on 24 October 2019 beginning on 08:52:39 UTC (ObsID 21465), with a net exposure of 43.60 ks.  A 1/8 sub-array was employed on the ACIS-S3 chip, reducing the nominal 3.2~s frame time by a factor of 8.0 to just 0.4~s, in order to limit photon pile-up.  The data were reduced using CIAO version 4.12 and the associated CALDB files.  In addition to the nuclear source, a moderately bright source is offset from the nucleus by approximately 5'' (see Figure \ref{fig:chandra_img}).  The position of the nuclear source derived in the Chandra source catalog version 2.0 (Evans et al.\ 2010) is 13:13:27.47, $+$36:35:38.20 (J2000), with an uncertainty of 0.7''.  We note that the center of the error region is nominally 1.1'' from the dynamical center of the CO velocity field determined by Kohno et al.\ (2003), but this may not be significant given the uncertainty in the X-ray position.  The position of the off-nuclear source is 13:13:27.10, +36:35:39.13, with an uncertainty of 0.9''.  We ran the CIAO tool \texttt{wavdetect} in order to determine the optimal extraction regions for each source.  These regions were consistent with circles with radii of 1'', so we used these simple regions to extract source counts.  Background regions were subtracted from adjacent regions, avoiding overlap between these sources.

The spectral and response files for the central source and its neighbouring sources were created using the tool \texttt{specextract}, based on the source positions obtained from \texttt{wavdetect}. The resulting spectra were binned using ftool \texttt{ftgrouppha}; a signal-to-noise ratio of 5.0 was used in order to optimize the sensitivity whilst being able to clearly distinguish the line and the continuum flux.

The image shown in Figure \ref{fig:chandra_img} was created through sub-pixel event repositioning.  Essentially, because the charge cloud deposited by an incident X-ray is smaller than a single pixel, and the dither of the telescope moves the pixel grid relative to the source image, it is possible to sample the source image on scales smaller than the pixel scale and the PSF of the telescope (see, e.g., Li et al.\ 2004).  This technique has previously been employed to reveal sub-arc second structure in nearby AGN, even in specific energy bands (e.g., Fabbiano et al.\ 2017).  An effective minimum scale of approximately 0.3'' is set by the absolute pointing uncertainty of the spacecraft aspect (e.g. Miller et al.\ 2017).  We used CIAO tools to create sub-pixels with sides equal to 0.1 native pixels (0.049'' vs 0.49''), and then smoothed the sub-pixels by a factor of 3.0 using Gaussian smoothing.

\subsection{XMM-Newton}

The XMM-Newton data were obtained from the HEASARC public archive.  Observation 0094360501 was made on 18 December 2002, starting at 16:05:55 UTC.  Using SAS version 18.0.0 and the contemporaneous calibration files, we ran the standard EPIC pipeline tools to produce event files.  Owing to our focus on spectroscopy in the Fe K band, we restricted our analysis to the EPIC-pn camera.  The pn was run in ``Prime Full Window'' mode during this observation.  Examining background regions, we found a modest degree of soft proton flaring within the observation; the SAS tool \texttt{tabgtigen} was used to create an event file to exclude the flaring intervals.  After filtering with ``FLAG=0'' and ``PATTERN=0-4'', we extracted source and background spectra using PI values of 0--20,479 grouped by a factor of 5.  The source and background regions had radii of 24 arcseconds; the background region was selected close to the source.  This filtering procedure yielded a net exposure of just 9.02~ks, and an average net count rate of 2.29~c/s (0.3--10.0~keV).  Response files were generated using the tools \texttt{rmfgen} and \texttt{arfgen}.  Prior to fitting, the spectrum was also binned to have a signal-to-noise ratio of 5.0 using the ftool \texttt{ftgrouppha}.  

We note that observation 0112551301 is also available in the public archive, but it is so severely affected by proton flaring that we have excluded it from this analysis.

\section{Analysis and Results} 

\subsection{Optical Spectroscopy and Source Classification}

The Palomar spectra were analyzed using SPEX version 3.06.01 (Kaastra et al.\ 1996).  The data were initially fit over a broad band in line-free regions to establish the continuum.  Combinations of narrow and broad Gaussian emission lines were then added to that continuum in narrow regions centered on the prominent lines, in order to characterize the line properties.  A $\chi^{2}$ goodness-of-fit statistic was minimized in the fitting process, and errors on optical parameters reflect the value of the parameter on the boundaries of their $1\sigma$ confidence intervals.

Figure 1 shows the optical spectra of NGC 5033, in the 4500--7000\AA~ range.  Broad Balmer lines are clearly evident.  Modeling of the broad H$\alpha$ line gives FWHM$=3400\pm 50~{\rm km}~{\rm s}^{-1}$, similar to the values reported by Koss et al.\ (2017) using more advanced techniques.  Simple modeling of the broad H$\beta$ line gives FWHM$=5400\pm 300~{\rm km}~{\rm s}^{-1}$.  These widths are fully consistent with an origin in the ``broad line region'' (or, BLR), and signal that the central engine is not obscured in NGC 5033.  

A narrow component to the H$\beta$ line is only barely evident, and poorly constrained in formal fits.  A narrow component is more prominent within the H$\alpha$ complex, but still only comparable in strength to the broad lines.  According to the classification scheme of Osterbrock (1981), this marks NGC 5033 as a Seyfert-1.5, at least at the time of our observations.  A Seyfert-1.8 designation would require very weak broad Balmer lines, and a Seyfert-1.9 designation would require the absence of broad H$\beta$ emission.  Neither of these classifications are consistent with our data.  ''Unobscured'' Seyfert-2 AGN are marked by the absence of broad (FWHM $\geq 1000~{\rm km}~{\rm s}^{-1}$) Balmer lines (see, e.g., Panessa \& Bassani 2002); NGC 5033 is also not consistent with this sub-class.

\subsection{Phenomenological X-ray Spectroscopy}

All X-ray spectral modeling was performed using XSPEC v12.11.1 (Arnaud 1996).  Fits to the Chandra data were made across the 1.0-8.0~keV band.  Owing to its larger collecting area and slightly higher sensitivity, it was possible to fit the XMM-Newton spectrum over the 0.3-10.0~keV band.  The quality of the fits were measured using $\chi^{2}$ statistics as binning to signal-to-noise ratio of 5 guarantees at least 25 counts per bin, in agreement with the recommendations of Cash (1979).  Initial fits with each spectral model were made using ``standard'' weighting, and then refined with the adoption of ``model'' weighting.  This has the effect of producing an improved fit to the continuum by not emphasizing falsely large errors toward zero flux.  All error values quoted throughout this paper represent 1$\sigma$ uncertainties.

We initially considered a simple phenomenological model consisting of a power-law continuum, and Gaussian line component to describe the Fe~K$\alpha$ line at 6.4~keV, \texttt{powerlaw+zgauss} in XSPEC terminology.  The redshift parameter within \texttt{zgauss} was set to that of NGC 5033 ($z=0.0029$). The results of this fit are given in Table \ref{table:powerlaw}.  Even with this simple model, a formally acceptable fit to both spectra is obtained.  The power-law indices are slightly harder than the median value of $\Gamma = 1.8$ for Seyferts (e.g., Nandra et al. 2007) but are within the range observed in the class.  The Fe~K$\alpha$ line energy values nominally differ between the Chandra and XMM-Newton spectra: the Chandra value of $E = 6.36^{+0.02}_{-0.02}$~keV is nominally redshifted, whereas the $E=6.43^{+0.02}_{-0.02}$~keV value for XMM-Newton is consistent with a small blue-shift or moderate ionization.  We note that the errors are merely $1\sigma$ intervals and that broader error bounds overlap.  We therefore regard the difference as not significant, and conclude that the line is consistent with largely neutral gas.  In both spectra, the line is unresolved.

\begin{figure*}[h]
\centering
\includegraphics[angle=-90,scale=0.5]{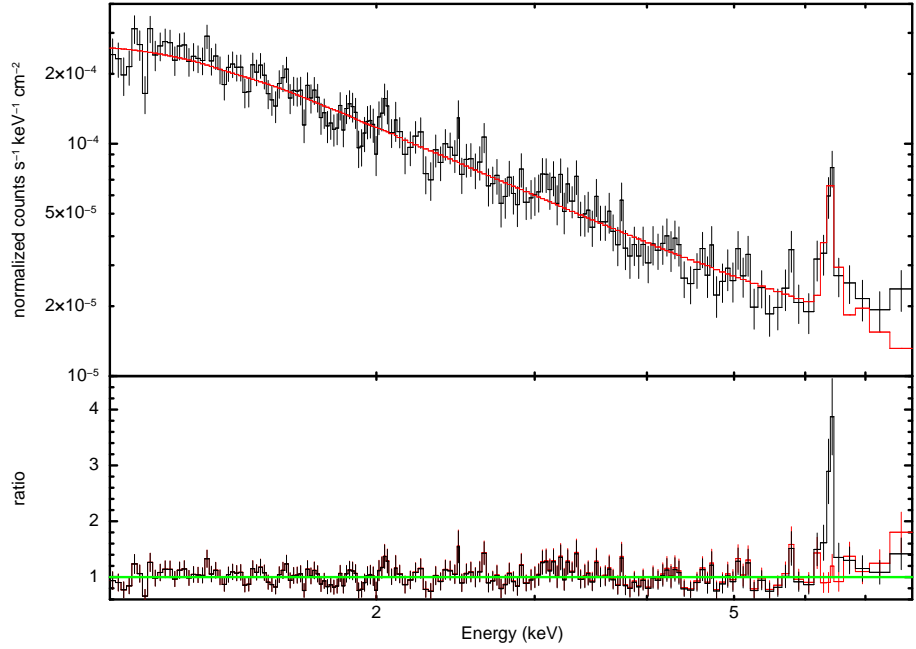}
\vspace {1mm}
\caption{The Chandra/ACIS spectrum of NGC 5033 obtained in 2019.  The model shown in red includes neutral reflection via the \texttt{pexmon} model.  The data/model ratio in red is obtained when this model is fit to the data; the ratio in black is that obtained when the Fe abundance within \texttt{pexmon} is set to zero to approximate a simple power-law fit.  The peak of the narrow Fe~K$\alpha$ line at 6.4~keV is nearly four times as strong as the local continuum.}
\label{fig:chandra}
\end{figure*}

\begin{table}[h]
\caption{Power Law with Gaussian Line}
\begin{footnotesize}
\begin{center}
\begin{tabular}{lll}
Parameter  &  Chandra   &  XMM-Newton \\
\tableline
$\Gamma$   &   $1.55^{+0.03}_{-0.03}$  &  $1.64^{+0.01}_{-0.01}$ \\
Norm. ($10^{-3})$ &   $0.319^{+0.008}_{-0.008}$ &  $1.064^{+0.008}_{-0.008}$ \\
\tableline
E (keV) &   $6.36^{+0.02}_{-0.02}$  &  $6.43^{+0.02}_{-0.02}$ \\
$\sigma$ ($10^{-2}$ keV) & $4.9^{+4.80}_{-4.90}$ & $3.7^{+3.9}_{-3.8}$ \\
Norm. $(10^{-5})$ & $0.8^{+0.2}_{-0.2}$ &  $1.3^{+0.3}_{-0.2}$ \\
EW (eV) &  $460^{+100}_{-90}$ & $250^{+40}_{-40}$ \\
\tableline
Flux $(10^{-11})$ & $0.263^{+0.007}_{-0.006}$ & $0.792^{+0.005}_{-0.005}$  \\
Luminosity $(10^{41})$ & 0.83(2) & 2.50(2) \\
\tableline
$\chi^{2}/\nu$ &   153.39/161 &   461.61/463 \\
\tableline
\end{tabular}
\vspace*{\baselineskip}~\\
\end{center} 
\tablecomments{The results of fits to the spectra of the nuclear black hole in NGC 5033 with a simple Gaussian plus power-law model.  The quoted errors are $1\sigma$ uncertainties. The equivalent width was measured using the \texttt{eqwidth} command in XSPEC.The continuum flux is quoted in units of ${\rm erg}~{\rm cm}^{-2}~{\rm s}^{-1}$ in the 0.3--10.0 keV~ band and the normalization is quoted in the units of $\rm{photons}~\rm{keV}^{-1}~\rm{cm}^{-2}~\rm{s}^{-1}$.}
\end{footnotesize}
\label{table:powerlaw}
\end{table}

The difference in the narrow Fe~K$\alpha$ equivalent width between the XMM-Newton observation in 2002 and the Chandra observation in 2019 is significant.  It is likely that this is dominated by changes in the continuum flux, which was three times higher in 2002 (see Table \ref{table:powerlaw}).  In contrast, while the line flux was nominally 60\% higher in 2002, this potential variation is small compared to the changes in the continuum.  It must also be noted that the errors on the line flux normalizations are large in a fractional sense, and that the $1\sigma$ confidence intervals overlap.

\subsection{Physical X-ray Spectral Modeling}

In view of the phenomenological results, we fit the spectra with a model that allows for intrinsic neutral absorption, and describes the Fe~K$\alpha$ line through X-ray reflection.  In XSPEC terminology, the improved model is: \texttt{ztbabs*(cutoffpl + pexmon)}. Here, the \texttt{ztbabs} component allows for neutral obscuration, and we fixed the redshift parameter to that of NGC 5033 ($z=0.0029$).   The \texttt{pexmon} component (Nandra et al.\ 2007) describes the reaction of a neutral slab of gas to an incident X-ray power-law spectrum with a high energy cut-off.  In our fits, \texttt{pexmon} replaces the Gaussian component.  In order to separate the reflected and direct emission, we fit \texttt{pexmon} as a pure reflection model by restricting the ``reflection fraction" to have negative values. The reflection fraction parameter traces the geometry of the reflector via $R = \Omega/2\pi$.  For consistency, the photon index, cut-off energy, and flux normalization of \texttt{pexmon} were linked to those of the cut-off power-law.  The fits are not sensitive to the reflector inclination nor the high-energy cut-off, so we simplified our fits by assuming $\theta = 60^{\circ}$ (the average viewing angle of a source in three dimensions) and $E_{cut} = 100$~keV as per Fabian et al.\ (2015).   We note that the inclination of an unobscured source should likely be lower (see, however, Elitzur et al.\ 2012), but in the absence of relativistic blurring, the predicted spectrum does not depend strongly on the inclination.
The Fe abundance relative to solar is degenerate with the reflection fraction in data of this quality; moreover, extraordinary supernova activity would be required to account for an elevated Fe abundance.  We therefore fixed all metal abundances to solar values within \texttt{pexmon}.

\begin{table}[ht]
\caption{Reflection Fits}
\begin{footnotesize}
\begin{center}
\begin{tabular}{lll}
Parameter  &  Chandra   &  XMM-Newton \\
\tableline
$N_H$ ($10^{22} cm^{-2}$)  &   $0.26^{+0.06}_{-0.06}$  &  $0.022^{+0.004}_{-0.004}$ \\
Redshift ($10^{-3}$)  &   2.92 &  2.92 \\
\tableline
$\Gamma$ &   $1.93^{+0.10}_{-0.09}$  &  $1.75^{+0.03}_{-0.02}$ \\
HighECut (keV)  & 100 & 100 \\
Norm. $(10^{-3})$ & $0.48^{+0.05}_{-0.04}$ &  $1.17^{+0.02}_{-0.02}$ \\
\tableline
Rel\_refl  & $-4.1^{+0.9}_{-1.2}$ &  $-1.9^{+0.3}_{-0.4}$ \\
Abundance   & 1.00  &  1.00 \\
Fe abundance   & 1.00  &  1.00 \\
Incl   & 60.0  &  60.0 \\
\tableline
Flux $(10^{-11})$ & $0.23^{+0.02}_{-0.02}$ & $0.78^{+0.01}_{-0.01}$  \\
Luminosity $(10^{41})$ & 0.73(7) & 2.50(3) \\
\tableline
$\chi^{2}/\nu$ &   137.92/157 &   439.9/464 \\
\tableline
\end{tabular}
\vspace*{\baselineskip}~\\
\end{center} 
\tablecomments{The results of fits to the spectra of the nuclear black hole in NGC 5033 with a \texttt{zphabs*(cutoffpl+pexmon)} model.  The quoted errors are $1\sigma$ uncertainties. The ``reflection fraction'' is negative to indicate that \texttt{pexmon} was fit in a reflection--only mode; a reflection fraction of 1.0 corresponds to a structure that covers half of the sky ($2\pi$ SR) as seen from the source, such as an accretion disk. The continuum flux is quoted in units of ${\rm erg}~{\rm cm}^{-2}~{\rm s}^{-1}$ in the 0.3--10.0 keV~ band and the normalization is quoted in the units of $\rm{photons}~\rm{keV}^{-1}~\rm{cm}^{-2}~\rm{s}^{-1}$.  The luminosity is quoted in units of ${\rm erg}~{\rm s}^{-1}$.  Model parameters without errors were fixed within the fit.}
\end{footnotesize}
\label{table:reflection}
\end{table}
\medskip

\begin{figure*}[t]
\centering
\includegraphics[angle=-90,scale=0.5]{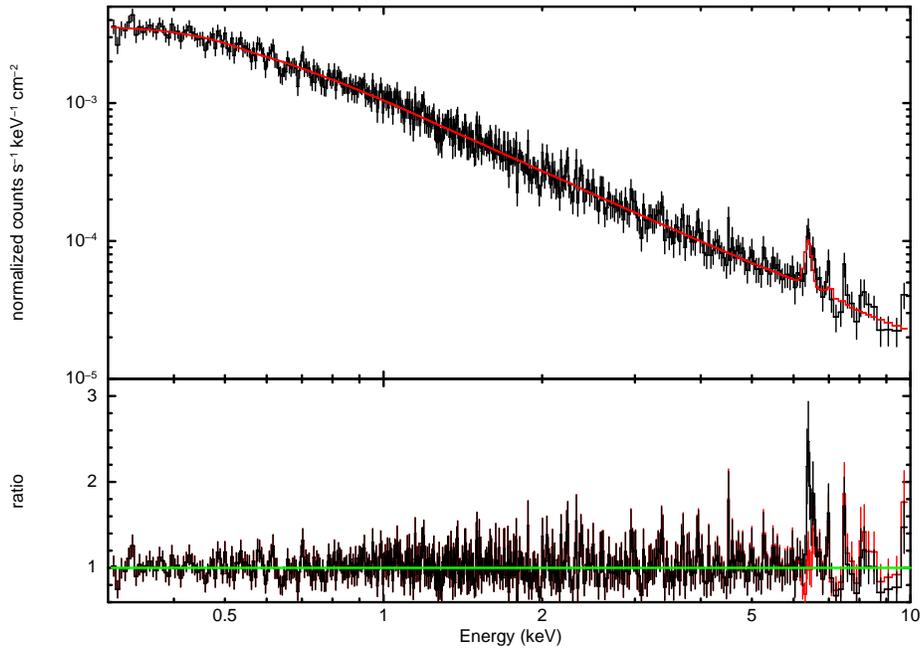}
\vspace {1mm}
\caption{The XMM-Newton/EPIC-pn spectrum of NGC 5033 obtained in 2002. The colors are the same as in Figure \ref{fig:chandra}. Here, the peak of the narrow Fe K$\alpha$ line is 2--3 times stronger than the local continuum.}
\label{fig:xmm}
\end{figure*}

The results of fits to this model, including intrinsic absorption and reflection, are listed in Table \ref{table:reflection}. Figures \ref{fig:chandra} and \ref{fig:xmm} show the fitted Chandra and XMM-Newton spectra.  Formally acceptable fits to both spectra were obtained using this physically motivated model, as indicated by the $\chi^2/\nu$ values of 137.92/157 and 439.9/464, respectively.  After separating the direct and reflected emission, the power-law indices become consistent with canonical values for Seyfert-1 AGN.  The absorption columns measured by this model are low, and inconsistent with an obscured central engine.  A higher column density is measured in the recent Chandra spectrum than in the archival XMM-Newton spectrum ($N_{H} = 2.6\pm 0.6 \times 10^{21}~{\rm cm}^{-2}$ and $N_{H} = 0.22\pm 0.04 \times 10^{21}~{\rm cm}^{-2}$, respectively).  The different values could be ascribed to variable obscuration, but it is more likely that differences in the fitting band and detectors account for the apparent variability.  The XMM-Newton data extend down to 0.3 keV whereas the Chandra data could only be considered above 1 keV owing to its intrinsically lower effective area (and degradation of this effective area over time).  A Galactic column density of $N_{H} = 1.0\times 10^{21}~{\rm cm}^{-2}$ is expected, based on the implementation of the HI4PI Survey (HI4PI Collaboration, 2016); this is consistent with the XMM-Newton measurement.

The most important results obtained with this model concern the strength of the reflected emission.  As viewed from the central engine, the inner disk covers a large fraction of the solid angle, but this geometry produces broad emission lines, while narrow lines are generally produced at much larger radii.  In the recent Chandra spectrum of NGC 5033, the reflection fraction is measured to be $R = 4.1^{+1.2}_{-0.9}$ (nominally unphysical), whereas it was measured to be $R = 1.9^{+0.4}_{-0.3}$ in the prior XMM-Newton spectrum.  

Results from the 70-month Swift/BAT survey of AGN find that approximately half of non-blazar AGN are obscured, when a column of $N_{H} = 1\times 10^{22}~{\rm cm}^{-2}$ is adopted as the dividing line (Ichikawa et al.\ 2019).  This is supported by H$\beta$ line diagnostics (Koss et al.\ 2017).  The fact that about half of Seyferts are obscured leads to the conclusion that the ``torus'' subtends about half of the sky (as seen from the central engine), or about $2\pi$ steradians ($R = \Omega/2\pi = 1$).   We note, however, that a larger fraction of low-luminosity or low-Eddington fraction sources may be obscured (e.g. Ricci et al.\ 2015).  

The spectra do not statistically require ionized absorption.  When ionized partial covering absorption is included in the models using the \texttt{zxipcf} component (see, e.g., Reeves et al.\ 2008), the fitting process forces the component to high ionization (log $\xi\simeq$ 5) and implausibly small covering factors ($f_{\rm cov} \sim 0.001$).  This has the effect of minimizing the lines predicted by \texttt{zxipcf}.  In order to obtain representative limits on the strength of any ionized absorption within the central engine, we made fits with plausible parameters (log $\xi = 3$ and $f_{cov} = 0.9$), fixed at the red-shift of the host galaxy.  We measure 90\% confidence upper limits of $N_{H} \leq 2\times 10^{21}~{\rm cm}^{-2}$ (Chandra) and $N_{H} \leq 6.5\times 10^{20}~{\rm cm}^{-2}$ (XMM-Newton).  We note that \texttt{zxipcf} samples a broad range of ionization very coarsely, and may not be suitable for detailed modeling of complex warm absorbers, but it permits meaningful limits at CCD resolution.

\subsection{The X-ray Baldwin Effect}

The unusually strong reflection in NGC 5033 can also be considered in the context of the X-ray Baldwin effect, also known as the Iwasawa-Taniguchi effect (Iwasawa \& Taniguchi 1993).  In short, the equivalent width of narrow Fe~K$\alpha$ lines may fall with increasing X-ray luminosity (and with increasing Eddington fraction), in analogy with the behavior of optical and UV lines in Seyferts and quasars.  The X-ray Baldwin effect, while not completely understood, is often interpreted as a decrease in covering fraction with increasing X-ray luminosity (e.g. Page et al.\ 2004, Bianchi et al.\ 2007).  In contrast, the optical and UV Baldwin effect is likely to be partially driven by photoionization and optical depth effects (e.g., Korista et al.\ 1998, Goad et al.\ 2004).

Shu et al.\ (2007) confirm evidence of the effect in X-rays, both in terms of raw X-ray luminosity and when using $L_{2-10}/L_{Edd}$ as a proxy for the mass accretion rate.  They find that their Chandra/HETG AGN sample is consistent with separate treatments of independent data by Page et al.\ (2004) and Bianchi et al.\ (2007), giving EW$\propto L^{-0.17\pm0.08}$ and EW$\propto (L_{bol}/L_{Edd})^{-0.19\pm0.05}$ (respectively).  We note that Shu et al.\ (2010) considered $L_{x}/L_{Edd})$ whereas Bianchi et al.\ (2007) considered $L_{bol}/L_{Edd}$, but the results are formally consistent.  

To place NGC 5033 within this context, we calculated 2--10 keV X-ray luminosity values using the fluxes reported in Table 2. We also took the X-ray luminosity and equivalent width from the 1999 ASCA observations (Terashima et al. 1999).  Koss et al.\ (2017) report a black hole mass of  $log ~\left(M_{BH} / M_{sun} \right) = 7.86\pm 0.35$ via the viral method (e.g., single-epoch H$\beta$ line fitting, following the methods in Trakhtenbrot \& Netzer 2012) and this value is used throughout our analysis.  The bolometric corrections were then estimated by analyzing the variation in correction factor with $L_{X}$ in the work of Vasudevan \& Fabian (2007).  We constructed bins in X-ray luminosity and used the average value of the correction factor in each bin to translate our X-ray luminosities into bolometric values.

\begin{figure*}[t]
\centering
\begin{minipage}{.47\textwidth}
    \centering
    \includegraphics[width=1.0\textwidth]{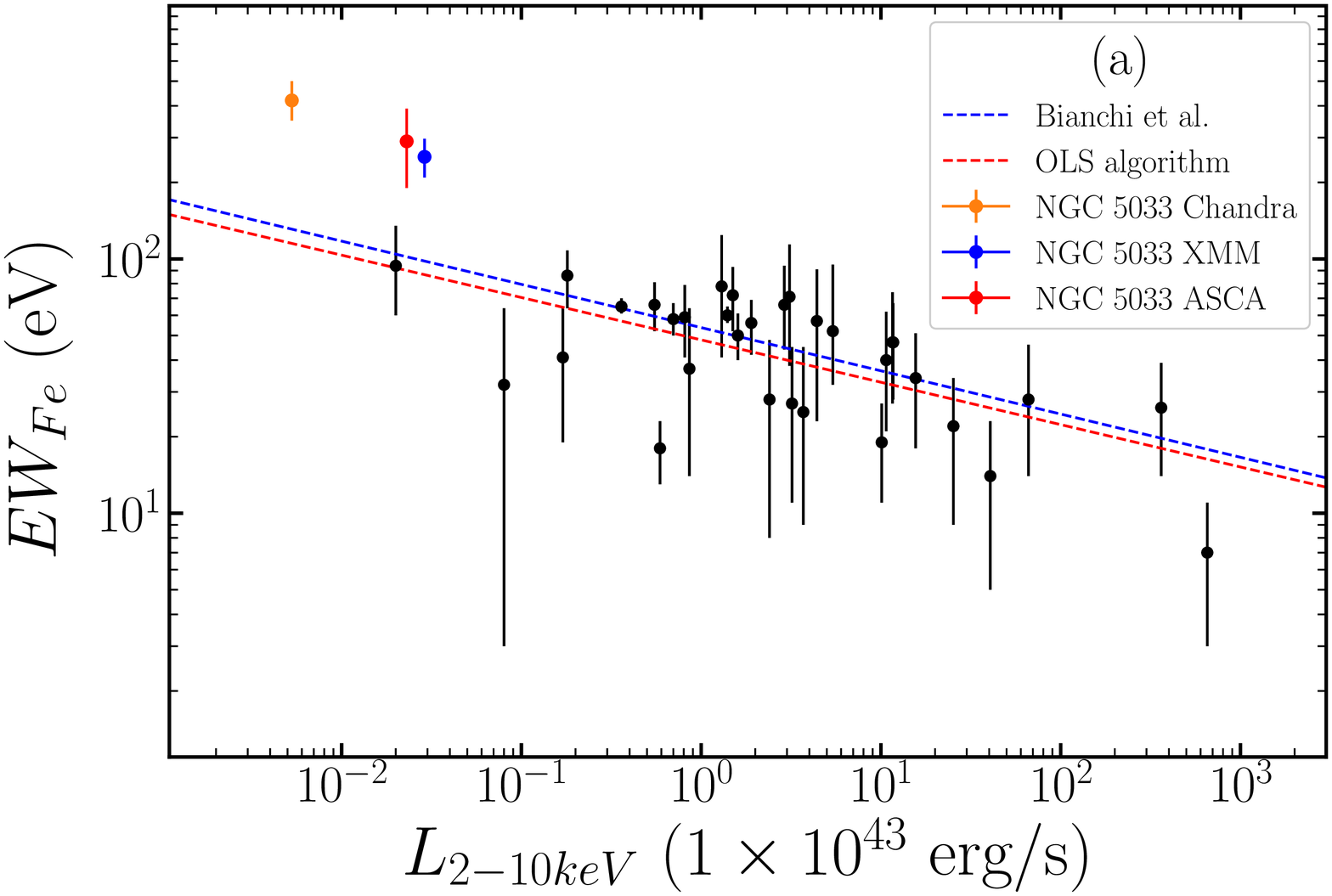}
\end{minipage}%
\hfill
\begin{minipage}{.47\textwidth}
    \centering
    \includegraphics[width=1.0\textwidth]{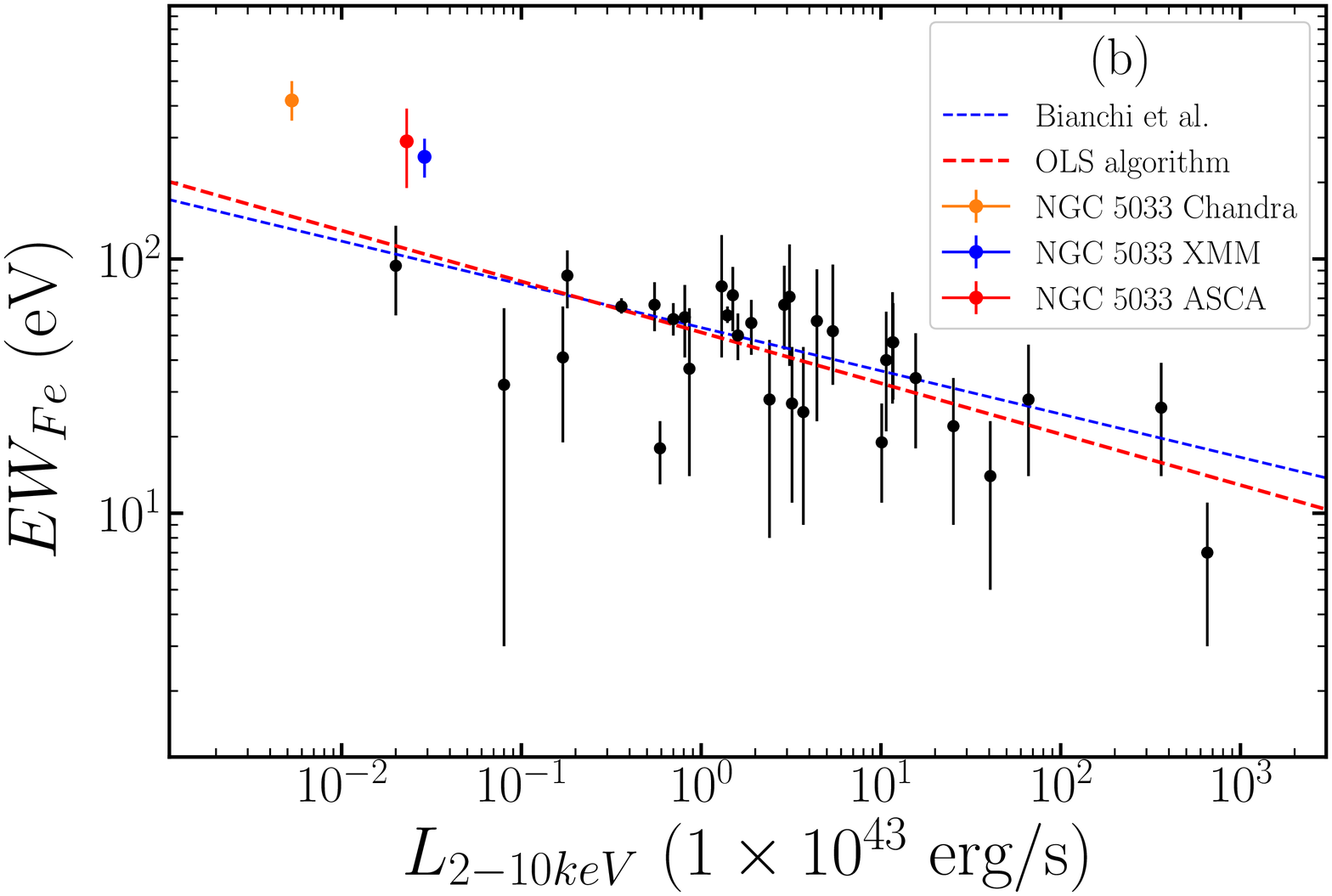}
\end{minipage}

\medskip
\medskip

\begin{minipage}{.47\textwidth}
    \centering
    \includegraphics[width=1.0\textwidth]{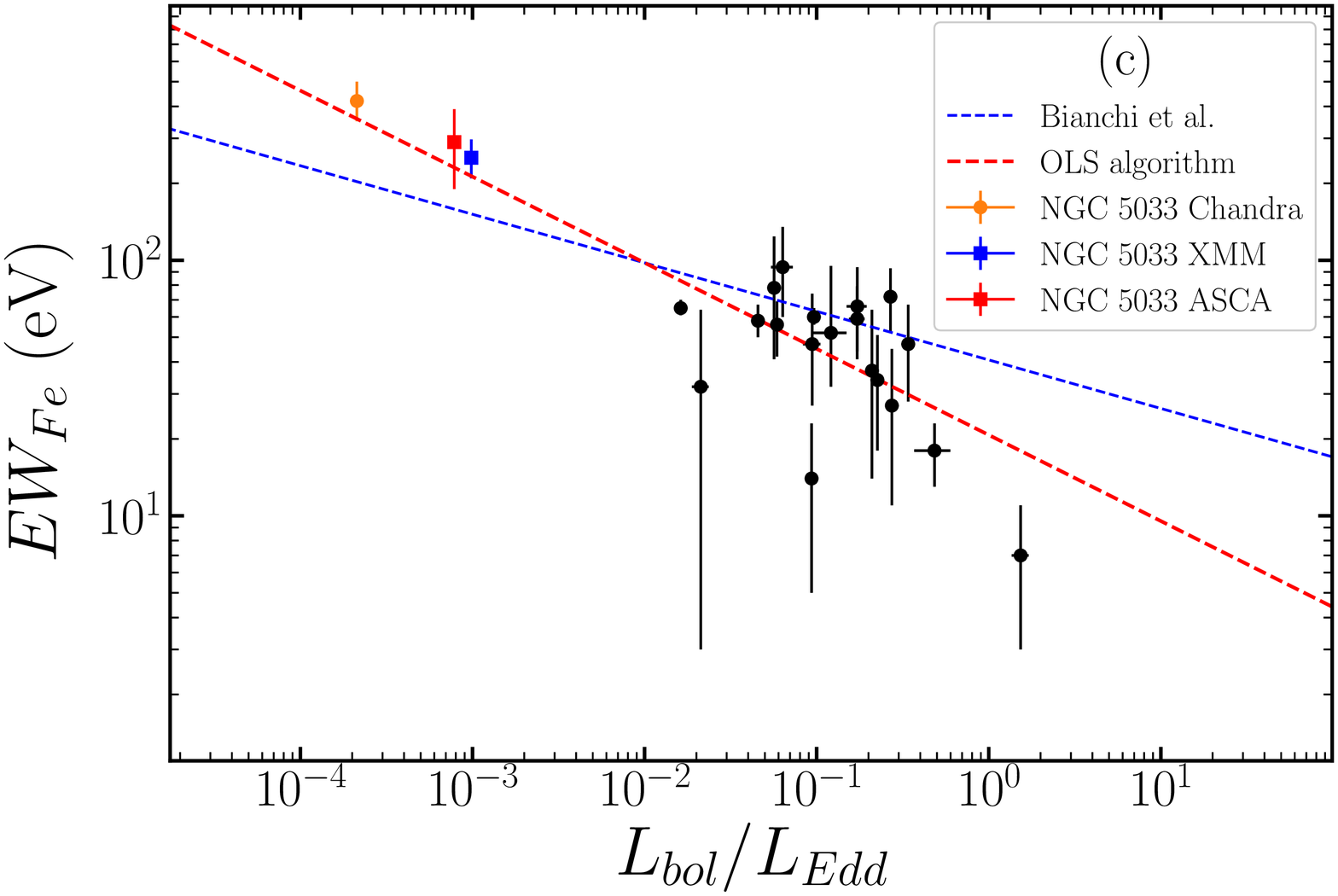}
\end{minipage}
\hfill
\begin{minipage}{.47\textwidth}
    \centering
    \includegraphics[width=1.0\textwidth]{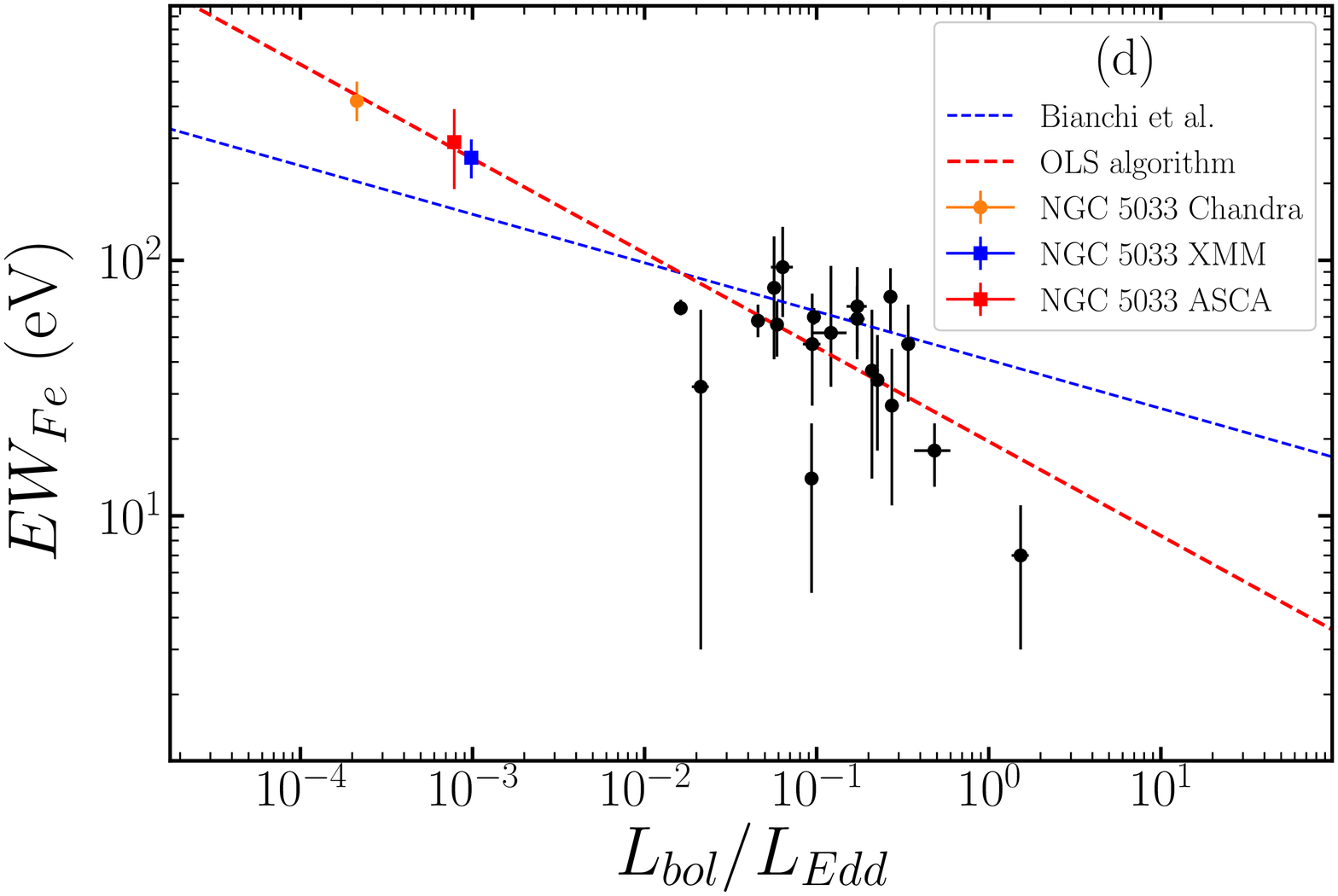} 
\end{minipage}
\caption{(a) EW of Fe K$\alpha$ emission line vs. the 2-10 keV luminosity, the red dashed lines are the linear fits calculated without including the points for NGC 5033 and the green dashed lines are the relation reported by Page et al. (2004). (b) As (a), but NGC 5033 was included in the linear fit. (c) EW of Fe K$\alpha$ emission line vs $L_{bol}/L_\text{Edd}$. The linear fits (red) were calculated without including NGC 5033. The green line is the relation reported by Bianchi et al. (2007). (d) As (c), but NGC 5033 was included in the linear fit.}
\label{fig:shu}
\end{figure*}

Figure \ref{fig:shu} shows the Fe~K$\alpha$ line equivalent width versus X-ray luminosity and Fe~K$\alpha$ line equivalent width versus $L_{bol}/L_{Edd}$, for NGC 5033 and the sources analyzed by Shu et al.\ (2010).  Average values for each source are plotted, and only sources with masses in Bentz \& Katz (2015) are included in the EW vs $L_{bol}/L_{Edd}$ plots.  (For clarity, independent points are plotted for NGC 5033, but its effect on the trend is calculated using the average of these data.)   NGC 5033 lies significantly above the EW vs $L_{X}$ anti-correlation measured by Page et al.\ (2004).  When NGC 5033 is included in a refined fit, the measured anti-correlation is slightly steeper and stronger, but NGC 5033 still lies significantly above the trend.  Similarly, the points from NGC 5033 lie significantly above the anti-correlation reported by Bianchi et al.\ (2007), which is otherwise a good description of the EW vs $L_{bol}/L_{Edd}$ data.   However, when NGC 5033 is included, the anti-correlation not only becomes steeper and stronger, but NGC 5033 lies on the refined anti-correlation, though it may be expected given the large dynamic range between NGC 5033 and the other data points. The results of our fits to these data are reported in Table \ref{table:shu_fits}.  A key finding is that when NGC 5033 is included in the analyses, the slopes of the best-fit relationships exclude zero at the $5\sigma$ level of confidence.  

\begin{table}[htb]
\caption{X-Ray Baldwin Effect Fit Parameters}
\begin{footnotesize}
\begin{center}
\begin{tabular}{lll}
Relation  &  EW vs. $L_x$   &  EW vs. $L_x/L_\text{Edd}$ \\
\tableline
Fit without NGC 5033   &   $-0.17 \pm 0.04$  &  $-0.34 \pm 0.12$ \\
Fit with NGC 5033  &   $-0.20 \pm 0.04$ &  $-0.37 \pm 0.08$ \\
\tableline \\
Bianchi et al. (2007)  &  $-0.17 \pm 0.03$ & $-0.19 \pm 0.05$ \\
\tableline
\end{tabular}
\vspace*{\baselineskip}~\\
\end{center} 
\tablecomments{The slope parameters of the linear fit to EW vs. $L_x$ or EW vs. $L_x/L_\text{Edd}$ on a logarithmic scale.}
\label{table:shu_fits}
\end{footnotesize}
\end{table}

\subsection{An off-nuclear source}

 Finally, as shown in Figure \ref{fig:chandra_img}, a source is evident about 5'' from the nuclear source. The source had a count rate of $2.28 \pm 0.23 \times 10^{-3}$ counts/s, which was very weak in comparison to the central source which had a count rate of $1.52 \pm 0.02 \times 10^{-1}$ counts/s. Hence, we were unable to constrain its spectrum due to its low count rate. We fitted a simple power-law using Cash statistics, and fixed the power-law index at $\Gamma = 2$. The resulting flux measured in the 1--8~keV band is $F = 4.1^{+0.5}_{-0.5} \ \times \ 10^{-14} \ \text{erg} \ \text{cm}^\text{-2} \ \text{s}^\text{-1}$.  At the distance of NGC 5033, this implies an X-ray luminosity of $L = 1.3\pm 0.2 \times 10^{39}~{\rm erg}~{\rm s}^{-1}$.  This is just above the threshold for ultra-luminous X-ray sources (ULXs), powered by Eddington-limited or super-Eddington accretion onto a neutron star or stellar-mass black hole.  Considering a circle with a radius of 15'', the X-ray background characterizations given by Brandt \& Hasinger (2005) suggest that there is only a 1.8\% chance of a background AGN with the same flux level within this region.

\section{Discussion}

New observations of NGC 5033 with Palomar and Chandra reveal an unobscured AGN.  Broad Balmer lines, comparable in strength to narrow components, clearly signal that the BLR is visible.  The X-ray spectrum is consistent with very little neutral obscuration, and does not require ionized absorption.  At least between 2019 and 2021, then, the source properties are consistent with a Seyfert-1.5 classification.  However, the Fe~K$\alpha$ emission line in the Chandra spectrum is several times stronger than is typical in Seyfert-1 AGN.  Past spectra obtained with XMM-Newton and ASCA also measured unusually strong, narrow Fe~K$\alpha$ lines.  In this section, we place our results into the broader context of unobscured and unobscured AGN, and suggest that a combination of flux lags and geometric considerations could explain both the unusual Fe~K$\alpha$ line and past discrepancies in the optical classification of NGC 5033.

In an analysis of 82 Chandra/HETGS observations of 36 unobscured Seyferts, Shu et al.\ (2010) find a mean narrow Fe K$\alpha$ line equivalent width of just EW$=53$~eV.  The values that we have measured in NGC 5033 (Chandra: $460^{+100}_{-90}$ eV , XMM-Newton: EW = $250^{+40}_{-40}$ eV ) are 8.7 and 4.7 times higher than this mean.  Indeed, these values are more consistent with the lines observed in obscured Seyfert-2 AGN.  In a companion analysis of 29 Chandra/HETGS observations of 10 obscured AGN, Shu et al.\ (2011) find a mean equivalent width of EW$=460$~eV; the remarkable Chandra value is consistent with this mean.  

Narrow Fe~K$\alpha$ lines arise through reflection, and reflection models can provide geometric constraints by measuring the relative importance of direct and reflected emission.  Within the \texttt{pexmon} model used in our work, the ``reflection fraction’’ is defined as $R = \Omega/2\pi$.  The value that we obtained in fits to the Chandra spectrum, $R = 4.1^{+1.2}_{-0.9}$, corresponds to more than $4\pi$~SR and is therefore unphysical.  The value obtained in our fits to an archival XMM-Newton spectrum is only barely below this limit ($R = 1.9^{+0.4}_{-0.3}$).    In contrast, in fits to 37 observations of 26 Seyferts, Nandra et al.\ (2007) report that only three spectra require a reflection fraction above unity.  The mean value of the Nandra sample is $R_{mean} = 0.51$, but the median value of $R_{median} = 0.39$ is likely more representative.  The results obtained from NGC 5033 are even more extreme when viewed in this context.  

Geometric inferences from reflection modeling are only valid when the central engine is steady, or when variations from the central engine and the response of the reflector (e.g., the molecular torus) can be tracked.  If a source is observed when the central engine flux is low but the reflector is still responding to flux that was higher in the past, faulty geometric inferences can result.   In a study of dust reverberation in quasars, Minezaki et al.\ (2019) developed a relationship between the size of the dusty torus and the luminosity of the central engine in AGN.  The relationship may not extend down to the luminosities inferred in NGC 5033, but if it does the luminosities that we infer from Chandra and XMM-Newton imply that the torus is just $\tau$=8–12 light-days from the central engine.  This is far shorter than the years between the observations made with XMM-Newton (2002) and Chandra (2019), so a degree of the unusual line flux in NGC 5033 can simply be ascribed to flux lags.

The continuum flux and Fe~K$\alpha$ line equivalent width observed with XMM-Newton and ASCA are very similar, and these values are still several times higher than those typically found in unobscured Seyferts.  While it is possible that a given observation suffers from a mismatch between the central engine and reflector, it is less likely that every observation caught a low continuum flux period.  This is particularly true when one considers that the light travel time between the central engine and even the torus in NGC 5033 is likely to be relatively short, as the periods of central engine flux enhancements and decrements then have to be fine-tuned.  It is here where past optical studies of NGC 5033 may provide additional clues.

Our observations with Palomar in 2021 demand a Seyfert-1.5 classification, based on the detection of broad Balmer lines consistent with an origin in the BLR.  The prior classification of NGC 5033 as a Seyfert-1.5 was also obtained using Palomar (Ho et al.\ 1997), potentially signaling that the sensitivity of this telescope is the key to detecting broad H$\beta$ emission.  However, it must be noted that Osterbrock \& Martel (1993) arrived at a Seyfert-1.9 classification based on observations with the 3m telescope at the Lick Observatory, and that the goal of their program was to search for weak, broad Balmer lines in the CfA Seyfert Sample (Huchra \& Burg 1992).  It is unlikely that a broad H$\beta$ line was simply overlooked in that investigation, and it is more likely that the optical spectrum of the source is variable.  This is consistent with variable obscuration within NGC 5033 that might also contribute to a heightened reflection fraction in X-rays.  

A similar scenario may be at work in NGC 4151, which is observed to pivot in flux and obscuration in X-rays, and to have an unusually strong narrow Fe~K$\alpha$ line (e.g., Miller et al.\ 2018).  Indeed, it is one of the rare ``unobscured’’ AGN wherein the reflection fraction is sometimes found to exceed unity (e.g., Nandra et al.\ 2007). Osterbrock \& Martel (1993) note that ``NGC 4151 has been observed to have variable Balmer lines over the past 20 years, being observed mostly as a Seyfert 1.5, but occasionally as a Seyfert 1.8 or 1.9 (Cohen \& Antonucci 1983, Penston \& Perez 1984).’’   The behavior inferred in NGC 5033 is certainly consistent with these changes.

In ``changing look’’ AGN,  changes in the inner accretion flow can precipitate changes in line fluxes and ratios that cause the formal classification to switch (e.g., Shappee et al.\ 2014, Yang et al.\ 2018).  Sources like NGC 4151 are sometimes discussed as a changing-look AGN, although the mechanism may be tied to a relatively high accretion rate (e.g., Sniegowska et al. 2020).  The fact that NGC 5033 displays similar changes signals that the changing-look phenomenon, at least as manifested in swings between unobscured sub-classes, is not exclusively tied to a high Eddington fraction.  The changes in NGC 4151 may alternatively be attributed to an accelerating outflow, launched close to the black hole (Shapovalova et al.\ 2010).  This may be a more plausible explanation for the optical variations seen in NGC 5033.  The outflow could also contribute to X-ray reflection, or serve to sometimes screen the reflector from the central engine. 

Regardless of whether the most extreme reflection in NGC 5033 can be explained through lags, or a combination of lags and geometric changes, the fact of three observations that all point to strong reflection suggests that the baseline geometry may be different than in many other Seyferts.  Whereas the optical/UV Baldwin effect is driven by photoionization and optical depth effects, the X-ray Baldwin effect has been ascribed to changes in the reflection covering factor (Iwasawa \& Taniguchi 1993, Page et al.\ 2004, Bianchi et al.\ 2007).  We have found that NGC 5033 lies above prior relationships describing the X-ray Baldwin effect, and that NGC 5033 increases the significance of the trends if it is explicitly included.  Explanations for the variable reflection geometry include luminosity-dependent winds at the face of the torus that would serve to increase its covering factor (Konigl \& Kartje 1994).  

Obscured but luminous black holes have been shown to preferentially occur in the late stages of major mergers and the post merger phases, agreeing with the theoretical expectations that merger-triggered accreting black holes are likely to be more luminous than those growing by slower secular processes (Koss et al.\ 2018). Using the Swift/BAT sample of AGN, Ricci et al.\ (2017) confirm that radiation pressure from the accretion disk is the main physical mechanism regulating the distribution of circumnuclear material on parsec scales.  Optical integral field unit spectroscopy of nearby AGN confirms that winds convey feedback onto larger scales (e.g., Harrison et al.\ 2014), broadly confirming seminal theoretical work showing that AGN winds can shape post-merger galaxies (e.g., Di Matteo et al.\ 2005).  At the Eddington fraction inferred in NGC 5033, the covering factor of obscuring (reflecting) material should be close to unity; our data suggest that the baseline value of reflection in NGC 5033 could be this high.  

However, it is unlikely that the black hole in NGC 5033 is clearing its environment in a manner that follows from this evolutionary scheme, owing to its modest Eddington fraction.  A more plausible explanation of its unusual reflection fraction may come from a study of the Circinus galaxy.  Ohsuga \& Umemura (2001) proposed that relatively low-luminosity AGN may be surrounded by relatively large walls of surrounding dust.  Such structures might be absent in more luminous AGN owing to the enhanced radiation pressure from the central engine.  It is also notable that distant reflection clouds have been found in the environment around Sgr A*, and in other nearby galaxies, (e.g., Ponti et al.\ 2010, Fabbiano et al. 2017, Marinucci et al. 2017, Jones et al. 2020, Yan et al. 2021).  This is a plausible explanation for a high baseline level of distant reflection in NGC 5033.  However, this mechanism may not operate below a certain threshold in Eddington fraction, as the LINER/LLAGN M81* does not show strong neutral Fe~K$\alpha$ emission lines: L$\sim10^{-5}~{\rm L}_{\rm Edd.}$, EW$=47\pm25$~eV (Young et al.\ 2007).

Our observations reveal that NGC 5033 is a source that may help to improve our understanding of AGN demographics, and the evolution of accretion flows onto massive black holes.  High-cadence, optical spectroscopic monitoring of NGC 5033 can help to understand the cadence at which its broad H$\beta$ emission varies.  Contemporaneous monitoring in X-rays can help to discern the relationship between changes in the optical appearance of NGC 5033, and the strength of the Fe~K$\alpha$ line, providing a broader view of changes in the accretion flow.  In the near future, high-resolution spectroscopy with XRISM (Tashiro et al.\ 2020) can measure the width of the narrow Fe~K$\alpha$ line in NGC 5033, clearly revealing its origin within the accretion flow.

We thank the Chandra X-ray Observatory for executing this observation.  The work of Daniel Stern was carried out at the Jet Propulsion Laboratory, California Institute of Technology, under a contract with NASA.  We thank the anonymous referees for constructive comments that improved this manuscript.





\end{document}